\documentclass[11pt]{iopart}

\usepackage{iopams}
\usepackage{amsfonts}
\usepackage{amssymb}
\usepackage{graphicx}
\usepackage[caption=false]{subfig} 
\usepackage{cite}

\usepackage[perpage]{footmisc} 

\expandafter\let\csname equation*\endcsname\relax
\expandafter\let\csname endequation*\endcsname\relax
\usepackage{amsmath}

\usepackage{bbm}
\newcommand{\id}{\mathbbm{1}}
\usepackage{bm}
\newcommand{\boldvec}[1]{\bm{#1}}

\def\colorfig{}

\newenvironment{acknowledgments}{\ack }{}

\usepackage{braket}%
\newcommand{\projector}[1]{\ket{#1}\bra{#1}}

\usepackage{hyperref}

\begin{document}
\title[Evaluation of bipartite entanglement between modes using translation symmetry]{
Evaluation of bipartite entanglement between two optical multi-mode systems using mode translation symmetry
}
\author{
Jun-Yi Wu 
and Holger F. Hofmann
}
\address{
Graduate School of Advanced Sciences of Matter, Hiroshima University, Kagamiyama 1-3-1, Higashi Hiroshima 739-8530, Japan
}
\ead{junyiwuphysics@gmail.com}

%


\begin{abstract}
Optical multi-mode systems provide large scale Hilbert spaces that can be accessed and controlled using single photon sources, linear optics and photon detection. Here, we consider the bipartite entanglement generated by coherently distributing $M$ photons in $M$ modes to two separate locations, where linear optics and photon detection is used to verify the non-classical correlations between the two $M$-mode systems. We show that the entangled state is symmetric under mode shift operations performed in the two systems and use this symmetry to derive correlations between photon number distributions detected after a discrete Fourier transform (DFT) of the modes.
The experimentally observable correlations can be explained by a simple and intuitive rule that relates  the sum of the output mode indices to the eigenvalue of the input state under the mode shift operation. Since the photon number operators after the DFT do not commute with the initial photon number operators, entanglement is necessary to achieve strong correlations in both the initial mode photon numbers and the photon numbers observed after the DFT. We can therefore derive entanglement witnesses based on the experimentally observable correlations in both photon number distributions, providing a practical criterion for the evaluation of large scale entanglement in optical multi-mode systems. Our method thus demonstrates how non-classical signatures in large scale optical quantum circuits can be accessed experimentally by choosing an appropriate combination of modes in which to detect the photon number distributions that characterize the quantum coherences of the state.

\end{abstract}

\maketitle

%
\section{Introduction}%
\label{sec::introduction}

The development of  large  scale  quantum  information  processing  in  quantum  optics systems \cite{EisenbergEtAlBouwmeester2004-BipartiteMultiPhEnt, NagataEtAlTakeuchi2007-4004, MatthewsEtAlOBrien2009-MPhCircuit, BogdanovEtAlTey2004-Qutrit, WieczorekEtAlWeinfurter2009-6PhDicke, AfekAmbarSilberberg2010-HighNooN, WangEtAlCleland2011-NOON, XiangHofmannPryde2013, IsraelEtAlSilberg2012-NOONTomography} can be achieved by combining the non-classical optical fields generated by single photon sources \cite{TakeuchiOkamotoSasaki2004-SinglePh, URenEtAlWalmsley2004-SinglePh, AtesEtAlSrinivasan2012-SinglePh, BernienEtAlHanson2012-SinglePhNVC, SanakaEtAlYamamoto2009-SinglePhSemicond, SantoriEtAlYamamoto2002-Singleph, KuhnHennrichRempe2002-SinglePh, KellerEtAlWalther2004-SinglePhIonTrap, MueckeEtAlRitter2013-SinglePhCavity, Eisaman2011-SinglePh, Takeuchi2014Review-SinPhEntPh} with increasingly complex networks of multi-mode interferometers \cite{ReckEtAlBertani1994-ExpSU(N),WeihsEtAlZeilinger1996-3MdMZIntfm, WalbornEtAlMonken2003-MltmdHOM, BarakBen-Aryeh2007-FTrnsQComByLinOps, PeruzzoEtAlOBrein2011-MltmdQIntf, SpagnoloEtAlSciarrino2012-3MdQIntfm, MetcalfEtAlWalmsley2013-MltmdQIntfmtry, ChaboyerEtAlSteel2015-3MdQIntf}.
It is now possible to realize large scale optical circuits by integrating optical fibers on chips or by writing the optical waveguides directly into a material by 3D printing. As a result, there has been an increasing interest in the possibilities of multi-mode interferences such as Boson sampling \cite{AaronsonArkhipov2011-CmplxLinOps, CrespiEtAlSciarrino2013-BsnSmplngOnChip, SpringEtAlWalmsley2013-BsnSmplOnChip, Ralph2013-QCmpBsnSmplOnChip}, multi-mode quantum metrology \cite{SpagnoloEtAlSciarrino2012-3MdQIntfm, HumphreysEtAlWalmsley2013-MltmdPhsEstm, ChaboyerEtAlSteel2015-3MdQIntf},
as well as new methods of entanglement generation between spatially separate multi-mode systems \cite{WisemanVaccaro2003-IdPtclEnt}.
Different from the well-studied case of two-path interferometers, the effects of a linear transformation of the modes on the photon statistics observed in the output of a multi-mode interferometer is much more difficult to characterize \cite{Scheel2004-PermanentInLO,TichyEtAlBuchleitner2010-MultiBS, SpagnoloEtAlOsellame2013-MltmdBsnBnch, CrespiEtAlSciarrino2016-SuppressionLawOfFT}.
In general, the output photon statistics of a well-defined photon number input can be determined by calculating the permanent of the corresponding transformation  matrix, but this is itself a NP-hard problem \cite{Valiant1979-PermSharpPHard, Scheel2004-PermanentInLO, Aaronson2011-LinOpPermSharpP}.
In the case of entanglement between two photonic multi-mode systems, it is therefore difficult to identify characteristic non-classical features that scale up naturally and remain easily accessible as the number of photons and modes increases.
As a result, it is desirable to develop more specific theories that make use of convenient properties of the interferometers, such as symmetries between the modes.

In this paper, we address the problem of identifying the characteristic signatures of the bipartite entanglement between two multi-mode systems that is naturally generated by passing the light emitted by single photon sources through beam splitters \cite{WisemanVaccaro2003-IdPtclEnt}.
When a large number of modes is split in this fashion and each of the output modes is distributed to two separate locations, the quantum state of the two multi-mode systems is highly entangled.
Although this entanglement is easy to generate experimentally, there will be a large number of imperfections and errors associated with the difficulties of controlling large numbers of modes and photons.
It is therefore necessary to identify experimentally observable criteria to judge wether a given experimental implementation is sufficiently close to the ideal entanglement predicted by theory. Specifically,  we need to formulate entanglement witnesses that can be applied to the measurement statistics of arbitrary quantum states, regardless of the types of errors and imperfections that may have occurred in the experiment.

In general, the experimental evaluation of entanglement requires the performance of separate measurements of non-commuting local properties in the two systems, e.g. the $X$ and $Z$ components of two entangled qubits. In multi-mode systems, it is possible to use the photon number distributions realized by different linear transformations of the modes. Specifically, the photon number operators of two complete sets of orthogonal modes do not commute with each other, and can therefore be used to construct entanglement witnesses.
The problem is that it takes complicated mathematical procedures to relate the linear transformation of modes to the resulting transformations of the Fock states representing multi-photon distributions.
It is therefore necessary to develop a simple and experimentally feasible strategy to identify the specific statistical signatures that would not be possible without entanglement.
For that purpose, we propose the application of discrete Fourier transforms (DFT) to relate two sets of modes to each other by multi-mode interferometry.
Although the DFT produces complicated photon number patterns by interfering all of the modes with each other, the symmetry of the operation under
mode shifts allows us to classify and characterize the patterns in a manner that enables us to efficiently identify the effects of  entanglement experimentally. Specifically, the mode shift symmetry of the DFT converts a cyclic shift of the input mode indices into a well-defined  phase shift of the output modes.
This useful property of the modes can be applied directly to the transformation of photon number states, resulting in a simple relation between the two representations of the quantum state which we call the mode shift rule of DFTs.
This mode shift rule identifies the quantum coherences between different input photon numbers of the DFT with sets of photon number distributions in the output, allowing us to identify the effects of quantum coherences in the entangled input state on the correlations between photon number patterns observed after the DFT has been applied to both multi-mode systems.
Based on this fundamental insight into the relation between photon number distributions before and after the DFT, we can then derive experimental criteria for the verification of entanglement between two multi-mode systems of arbitrary size and photon numbers.

The remainder of the paper is organized as follows. In section \ref{sec::ent_generation}, we discuss
the generation of entanglement between two local $M$-mode systems A and B
by beam splitting M single photon input modes. In section \ref{sec::partitions}, we analyze the distribution of photons between the systems A and B and consider the entanglement in each distribution subspace. In section \ref{sec::limit_lin_opt}, we consider measurements suitable for the verification of entanglement in the presence of experimental imperfections. We show that non-commuting measurements can be implemented by linear optics and photon detection and  identify the DFT as the most promising realization of non-commuting photon detection measurements . In section \ref{sec::DFT}, we formulate the mode shift rule of DFTs to describe the relation between quantum coherence in the input photon number basis and photon number distributions observed in the output of DFTs. In section \ref{sec::ent_crit}, we employ the mode shift rule to formulate an experimentally observable criterion
for entanglement verifications by identifying the bounds of correlation fidelities that
apply to all separable states of the two multi-mode systems. In section \ref{sec::tighter_bound}, we consider the statistics of pattern classes observed in the original input modes and derive an optimized entanglement witness. In section \ref{sec::examples},  we illustrate the evaluation procedure using the specific example of 4-mode single photon inputs split into two local 4-mode systems with 2 photons each.  Section \ref{sec::conclusions} concludes the paper.

\section{Entanglement from multi-mode beam splitting}
\label{sec::ent_generation}

Multi-mode systems can be characterized experimentally by photon number detection. Ideally, it is possible to assign a specific photon number $n_{m}$ to each mode $m$, so that the observed photon number distribution can be given by an array
\begin{equation}
  \boldvec{n} = (n_{0},...,n_{M-1}),
\end{equation}
where the modes are indexed by $\{0,...,M-1\}$.
Since the photon numbers are eigenvalues of the $M$ harmonic oscillator modes, each photon number distribution corresponds to a well defined energy eigenstate $\ket{\boldvec{n}}$ in the Fock space of the $M$-mode system.

Entangled states of two $M$-mode systems can be described by a  superposition of photon number states $\ket{\boldvec{n}_{A}}\ket{\boldvec{n}_{B}}$, where the photon number patterns $\boldvec{n}_{A}$ are correlated with the photon number patterns $\boldvec{n}_{B}$.
The simplest example is a single photon passing through a beam splitter, which results in a coherent superposition of the photon number state $\ket{1_{A}}\ket{0_{B}}$ where the photon is found in $A$, and the photon number state $\ket{0_{A}}\ket{1_{B}}$ where the photon is found in $B$. Technically, this superposition is a maximally entangled Bell state of the local modes at $A$ and at $B$ given by $( \ket{1_{A}}\ket{0_{B}} + \ket{0_{A}}\ket{1_{B}})/\sqrt{2}$ \cite{TanWallsCollet1991-SinglePhNL, Hardy1994-SinglePhNL, HessmoUsachevEtAlBjork2004-ExpSnglPhNnlcl,Enk2005-SglPtclEnt, Enk2006-Rpl}. However, photon detection can only access the $\{\ket{0},\ket{1} \}$ basis, making it impossible to observe the entanglement without additional ancillary photon sources.

\begin{figure}[t]
  \centering
  \includegraphics[width=0.5\textwidth]{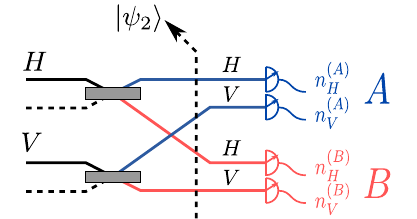}
  \caption{\colorfig
  The schematic setup of entanglement generation between two $2$-mode systems in \cite{OuMandel1988-TwoPhBiEnt}. Two single photons in the $\{H,V\}$ polarization modes are generated from a down conversion source, and distributed by beam splitters into two local system with the modes $\{H_{A},V_{A}\}$ and $\{H_{B},V_{B}\}$, respectively. The state $\ket{\psi_{2}}$ after the beam splitting is given in Eq. \eqref{eq::psi_2}. By post-selecting one-photon detection events in each local system, one projects the $\ket{\psi_{2}}$ onto the entangled state $\ket{\phi_{1_{A},1_{B}}}$ in Eq. \eqref{eq::phi_1A1B}.
  }
  \label{fig::ent_generation_2Ph}
\end{figure}%

Conveniently, the situation changes when this system is scaled up by combining multiple single photon inputs, distributing them to the multi-mode systems $A$ and $B$.
A well known example is the generation of entangled pairs used in one of the earliest experiments of entanglement generation by Ou and Mandel \cite{OuMandel1988-TwoPhBiEnt}, where two orthogonal modes are split by two beam splitters and the $4$ output modes are separated into two local systems with two modes each (see Fig. \ref{fig::ent_generation_2Ph}) .
Two single photons are generated in the modes $\{H,V\}$ and then distributed into the local systems $A$ and $B$ by the beam splitters, resulting in the state
\begin{equation}
\label{eq::psi_2}
  \ket{\psi_{2}} = \frac{1}{2}(\ket{00}_{A}\ket{11}_{B} + \ket{11}_{A}\ket{00}_{B}
  + \ket{01}_{A}\ket{10}_{B} + \ket{10}_{A}\ket{01}_{B}).
\end{equation}
By post-selection of the outcomes with only one photon in each local system , one can obtain the two photon Bell state
\begin{equation}
\label{eq::phi_1A1B}
  \ket{\phi_{1_{A},1_{B}}} = \frac{1}{\sqrt{2}}(\ket{01}_{A}\ket{10}_{B} + \ket{10}_{A}\ket{01}_{B}).
\end{equation}
As is well known, it is easy to access the entanglement of this state using linear optics and photon detection, since the two-mode transformations of linear optics map directly onto the Bloch sphere of the two-level single photon systems.

\begin{figure}[t]
  \centering
  \vspace{1em}
  \includegraphics[width=0.7\textwidth]{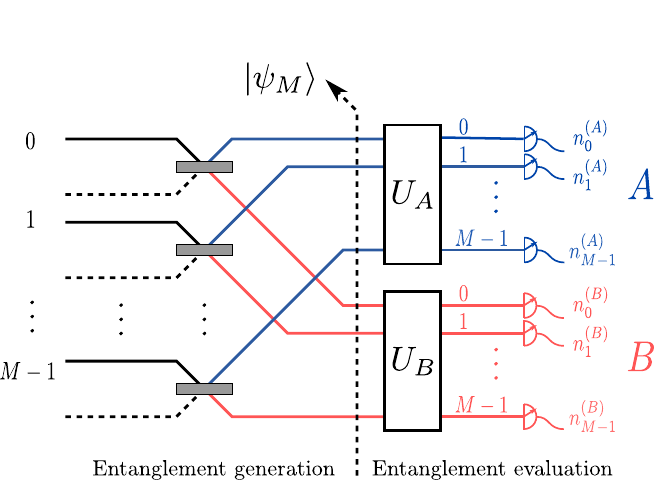}
  \caption{\colorfig
    Generation and evaluation of bipartite entanglement from $M$ single photon inputs. The input modes and the modes in each output port are indexed by $0,...,M-1$. The $M$ single photon inputs are split by $M$ beam splitters, generating the output state $\ket{\psi}$.
    For the evaluation, a selection of linear optics transformations $\hat{U}_{A}$ and $\hat{U}_{B}$ are applied to the modes, transforming the photon number operators in the different modes into sets of non-commuting observables.
  }
  \label{fig::ent_generation_nPh}
\end{figure}%

It is a straightforward matter to scale up the procedure of generating bipartite entanglement by beam splitting to general $M\otimes M$-mode systems \cite{WisemanVaccaro2003-IdPtclEnt}. As shown in the left part of Fig. \ref{fig::ent_generation_nPh}, $M$ photons in $M$ input modes can be split by $M$ beam splitters and distributed into two separate locations $A$ and $B$. Each local system then has $M$ modes described by mode indices $m=0,...,M-1$ that correspond to the mode index of the input mode from which they originated.
Since the photon in input mode $m$ will either appear in the corresponding mode $m$ of port $A$ or in the corresponding mode $m$ of port $B$, the state $\ket{ \psi_{M}}$ obtained after the photons have been distributed to $A$ and $B$ is given by a coherent superposition of all possible photon number states $\ket{\boldvec{n}_{A}}\ket{\boldvec{n}_{B}}$ satisfying the complementary correlation $\boldvec{n}_{A} + \boldvec{n}_{B} = (1,...,1)$,
\begin{align}
\label{eq::ent_generation_nPh}
  \ket{\psi_{M}}
  =
  \left(\frac{1}{\sqrt{2}}(\ket{1}_{A}\ket{0}_{B}+\ket{0}_{A}\ket{1}_{B})\right)^{\otimes M}
  = 
  \frac{1}{2^{M/2}}\sum_{n_{m}\le 1} \ket{\boldvec{n}}_{A} \ket{\bar{\boldvec{n}}}_{B},
\end{align}
where $\boldvec{n}$ are the $M$-dimensional vectors describing the photon number distribution in the $M$ modes. Each vector represents a specific distribution pattern, and in general, $n$ could take any value from $0$ to $M$. However, the input patterns are restricted to patterns with $n=0$ or $n=1$, so that it is possible to define the complementary pattern $\bar{\boldvec{n}}$ as $(1,...,1)-\boldvec{n}$. Consequently, the sum in Eq.(\ref{eq::ent_generation_nPh}) runs over all $M$-mode patterns with maximally one photon in each mode.

The state $\ket{ \psi_{M}}$ exists in a Hilbert space that includes many different photon number distributions between system $A$ and system $B$. However, linear optics and photon detection will always identify the precise value of the local photon number. It is therefore not possible to access the coherences between subspaces with different local photon numbers without using additional resources such as weak coherent input fields. For practical reasons, it is useful to distinguish the total photon numbers in the local systems before analyzing the entanglement that can be observed by detecting the precise distribution of photons within each local $M$-mode system.

\section{Distribution of photons between $A$ and $B$}
\label{sec::partitions}

Linear optics transformations always conserve the total photon number. This means that, no matter what linear optics transformation we chose, the total output photon number at the detectors remains the same. Conversely, it also means that linear optics and photon number detection is never sensitive to quantum coherences between states of different total photon number. We can therefore classify the measurements performed by linear optics and photon detection according to the total photon number $N$ detected in the output. If $\hat{\Pi}^{N}_{M}$ is the projector on the $M$-mode Fock states with a total photon number of $N$, we can express each measurement as a sequence of projection on total photon number, unitary transformation, and final detection of the specific photon number distribution $\ket{\boldvec{n}}$ with $|\boldvec{n}| = N$,
\begin{equation}
\label{eq::commuting_ph_num_proj_lin_optics}
  |\braket{\boldvec{n}|\hat{U}_{\text{lin.}}|\psi}|^{2} = |\braket{\boldvec{n}|\hat{U}_{\text{lin.}}\hat{\Pi}^{N}_{M}|\psi}|^{2}.
\end{equation}
Since the operator $\hat{\Pi}^{N}_{M}$ can be applied to the input state $\ket{\psi}$ before the linear optics mode transformation, it is sufficient to consider the quantum statistics of the reduced state $\hat{\Pi}^{N}_{M}\ket{\psi}$ by itself, separate from the components with different total photon numbers. In the case of the entangled state $\ket{\psi_M}$, we can apply the same reasoning to both local systems and make use of the fact that the total photon number is $M$. The probabilities for a specific set of detection patterns are then given by the probabilities
\begin{align}
\label{eq::commuting_ph_num_proj_lin_optics_1}
  P(\boldvec{n}_A,\boldvec{n}_B)
  = |\braket{\boldvec{n}_A\boldvec{n}_B|(\hat{U}_{\text{lin.}}^{(A)}\otimes\hat{U}_{\text{lin.}}^{(B)})(\hat{\Pi}^{N}_{M}\otimes\hat{\Pi}^{M-N}_{M})|\psi_M}|^{2},
\end{align}
where $N=|\boldvec{n}_{A}|$ is the total photon number detected in $A$ and $M-N=|\boldvec{n}_{B}|$ is the total photon number detected in $B$. Since any photon detection measurement clearly distinguishes between the different partitions of photon numbers between $A$ and $B$, it is useful to characterize the quantum state components associated with each partition separately.

The probability of obtaining a photon number partition of $(N,M-N)$ is given directly by the expectation value of the projectors and is easily obtained from the binomial statistics introduced by the beam splitters that distribute the photons to $A$ or $B$,
\begin{equation}
\label{eq::projection_prob}
  P(N,M-N) = \braket{\psi_{M}|\hat{\Pi}_{M}^{N}\otimes\hat{\Pi}_{M}^{M-N}|\psi_{M}} = \frac{1}{2^M} \binom{M}{N}.
\end{equation}
If we use the total photon number as a fixed condition in the evaluation of the entanglement, we can convert the measurement probabilities $P(\boldvec{n}_A,\boldvec{n}_B)$ into conditional probabilities by dividing them by $P(N,M-N)$. The measurement probabilities for specific patterns $(\boldvec{n}_A,\boldvec{n}_B)$ with $|\boldvec{n}_{A}|=N$ and $|\boldvec{n}_{B}|=(M-N)$ are then obtained directly from the normalized quantum state in the $(N,M-N)$ subspace, which is given by
\begin{equation}
\label{eq::projection_psi_M}
  \ket{\phi_{N,M-N}} =  \frac{1}{\sqrt{P(N,M-N)}}\hat{\Pi}_{M}^{N}\otimes\hat{\Pi}_{M}^{M-N}\ket{\psi_{M}}.
\end{equation}
Effectively, we can always post-select the state $\ket{\phi_{N,M-N}}$ in any measurement that identifies the total photon number in $A$ and in $B$. We can therefore simplify our analysis by looking at the entanglement of one post-selected state at a time. It should be noted that the state $\ket{\phi_{N,M-N}}$ is still an entangled state of two multi-photon $M$-mode systems. In fact, the post-selection does not change the structure of the entangled state much, as can be seen when it is expressed in the input mode photon number basis,
\begin{equation}
\label{eq::generated_ent_st_NANB}
  \ket{\phi_{N,M-N}} =
  \frac{1}{\sqrt{\binom{M}{N}}}
  \sum_{
    n_{m}\le1, |\boldvec{n}| =N
  } \ket{\boldvec{n}} \ket{\bar{\boldvec{n}}}.
\end{equation}
The entanglement is still expressed by perfect correlations between the photon number patterns in $A$ and their complementary patterns in $B$, and the only change with regard to the original state is that the sum runs only over patterns that have exactly $N$ photons in system $A$.
For example, with $4$ single-photon inputs one can generate the state $\ket{\psi_{M=4}}$ in the entanglement generation process of Fig. \ref{fig::ent_generation_nPh},
\begin{align}
\label{eq::psi_4M}
  \ket{\psi_{M=4}} =  
  \frac{1}{4} \left(\ket{\phi_{0_{A},4_{B}}}  + \vphantom{\sqrt{6}}
  2\ket{\phi_{1_{A},3_{B}}} + \sqrt{6}\ket{\phi_{2_{A}, 2_{B}}} + 2\ket{\phi_{3_{A},1_{B}}} +  \ket{\phi_{4_{A},0_{B}}}\right),
\end{align}
where $\ket{\psi_{4}}$ is a superposition of the components $\ket{\phi_{N,4-N}}$ with different photon number partitions.
The post-selection of the photon number partition $(2_{A},2_{B})$ from the $4$-photon state $\ket{\psi_{4}}$ is perfectly correlated in $6$ complimentary patterns,
\begin{align}
\label{eq::generated_ent_st_2A2B}
  \ket{\phi_{2_{A},2_{B}}} =  \frac{1}{\sqrt{6}}
  & \left(
    \ket{1100}\ket{0011} + \ket{0110}\ket{1001} + \ket{0011}\ket{1100} + \ket{1001}\ket{0110}
  \right. \nonumber
  \\
  & \left.
    + \ket{1010}\ket{0101} + \ket{0101}\ket{1010}
  \right).
\end{align}

The post-selected states are maximally entangled in a Hilbert space of dimension $\binom{M}{N}$, as compared to the original entanglement with a Schmidt rank of $2^M$. Specifically, the original $2^M$-dimensional entanglement is divided up into $M+1$ different components $\ket{\phi_{N,M-N}}$ of which all except the one-dimensional cases of $N=0$ and $N=M$ are entangled states. The two separable states $\ket{\phi_{0,M}}$ and $\ket{\phi_{M,0}}$ both have a probability of $1/2^{M}$, which falls off exponentially as the input photon number $M$ increases. In general, the Schmidt rank of all entangled states $\ket{\phi_{N,M-N}}$ is proportional to the probability of detecting the state $\ket{\phi_{N,M-N}}$ in the input $\ket{\psi_{M}}$. This means that the probability of detecting a state actually increases with the amount of entanglement in that state. The states with the highest probabilities $P(N,M-N)$ are the states with nearly equal photon numbers in $A$ and $B$. For high input photon numbers $M$, this probability can be approximated by a Gaussian distribution,
\begin{equation}
\label{eq::rankprob}
  P(N,M-N) \approx \sqrt{\frac{2}{\pi M}} \exp(- \frac{2}{M}(N-\frac{M}{2})^2).
\end{equation}
The Schmidt rank of each photon number partition is given by $2^M P(N,M-N)$, indicating that the vast majority of post-selected states have high amounts of entanglement. Since the Schmidt rank of the complete input state is $2^M$, the maximal probability of $P(M/2,M/2)$ is a good estimate of the expected reduction of Schmidt rank caused by the use of linear optics and photon detection.
As can be seen from Eq. \eqref{eq::rankprob}, this reduction ratio scales with the square root of $M$, which is nearly negligible when compared to the exponential scaling of the overall Schmidt rank. We can therefore conclude that the limitation to linear optics and photon detection reduces the amount of available entanglement only slightly while greatly reducing the technological effort involved in the characterization and application of the entangled state.
In the following, we will take a closer look at the possible measurements that can be performed to characterize the entangled state in experiments.

\section{Transformation of photon number states by linear optics}
\label{sec::limit_lin_opt}

In practice, it will be difficult to achieve precise control of the number of photons generated, and mode matching errors and losses may result in a wide variety of additional experimental imperfections.
It is therefore important to identify characteristic features of the entangled state that are robust against such imperfections, and can be used as experimental criteria for the verification of entanglement generation.
The basic procedure for entanglement evaluation is shown in right part of Fig. \ref{fig::ent_generation_nPh}.
Correlations between non-commuting physical properties can be observed by applying different linear optics transformations $\hat{U}_{A}$ and $\hat{U}_{B}$ to the local multi-mode systems.
These local unitaries change the measurement basis of photon detection by transforming the modes in which the photons are detected.
The transformed measurement basis corresponds to the photon number operators $\hat{n}'_{k}=\hat{U}^{\dagger}\hat{n}_{k}\hat{U}$ in the output modes $k=0,...,M-1$ of the local linear transformation $\hat{U}$, which do not commute with the original photon number operator $\{\hat{n}_{m}\}_{m=0,...,M-1}$ in the input modes.
It is therefore possible to verify entanglement by using only two pairs of unitaries, where one pair is given by the identity operation $\id\otimes\id$ and the other pair $\hat{U}_{\text{lin.}}\otimes\hat{U}_{\text{lin.}}$ generates a maximally complementary non-commuting photon number basis.
Here the application of $\hat{U}_{\text{lin.}}$ corresponds to the application of Hadamard gates to qubits in order to transform the computational $Z$-basis into the non-commuting $X$-basis.
It will then be possible to identify entanglement witnesses based on the probabilities of photon number distributions observed either in the input modes or after $\hat{U}_{\text{lin.}}$ has been performed.

We now need to find an optimal linear optics transformation to obtain a suitable complementary basis for the evaluation of the entanglement.
For this purpose, we need to consider the matrix elements that connect the output photon number distribution in output modes $k$ to the input photon number distributions in input modes $m$.
However, linear optics transformations are usually expressed by a unitary $M\times M$ matrix that described by the transformations of the $M$ creation operators associated with the coherent field amplitudes of the modes,
\begin{equation}
\label{eq::linopt}
  \hat{U}_{\mathrm{lin.}}^\dagger \hat{a}_{k}^{\dagger} \hat{U}_{\mathrm{lin.}}
  = \sum_{m=0}^{M-1} T^*_{km} \hat{a}_{m}^{\dagger}.
\end{equation}
In terms of photon number distributions, the coefficients $T_{km}$ represent the possibility of a photon transfer from mode $m$ to mode $k$.
It is in principle possible to construct all multi-photon matrix elements of $\hat{U}_{\text{lin.}}$ from Eq. \eqref{eq::linopt}.
However, the calculations involve complicated quantum interference effects, making it a non-trivial task to determine the complete multi-photon statistics of linear optics transformations.
Specifically, quantum interference results in photon bunching effects, which results in a preference for output patterns with multiple photons in the same mode. We can therefore conclude that any linear optics transformation applied to the entangled state $\ket{\phi_{N,M-N}}$ of Eq. \eqref{eq::generated_ent_st_NANB} which has at most one photon in each input mode is likely to result in output photon number distributions with more than one photon in each mode.
This means that most output measurements cover the complete Hilbert spaces of $N$ photons in $M$ modes, so that it is necessary to find efficient classification schemes in order to identify correlations between the different kinds of photon number distributions observed in system $A$ and in system $B$.

Since all of the modes are treated equally in the entanglement generation procedure, the optimal transformation should also be symmetric in the modes.
We should therefore consider a linear optics transformation that interferes equal fractions of each input mode with each other, resulting in a unitary matrix with $|T_{km}|^2=1/M$.
In addition, we need a transformation that is easy to apply to any number of modes. By combining these two requirements, we arrive at the discrete Fourier transform (DFT) $\hat{U}_{\mathrm{F}}$, given by
\begin{equation}
\label{eq::DFT_transformation}
  \hat{U}_{\mathrm{F}}^\dagger \hat{a}_{k}^{\dagger}\hat{U}_{\mathrm{F}}
  = \sum_{m=0}^{M-1} \frac{1}{\sqrt{M}}e^{-i\frac{2\pi}{M}km} \hat{a}_{m}^{\dagger}.
\end{equation}

For single photon states, the condition that all modes interfere equally corresponds to the requirement that the input photon number basis is mapped onto a mutually unbiased output basis. However, the results for multi-photon input states involve quantum interferences and photon bunching effects that result in a more complicated relation between the input basis and the output basis. In the following, we will simplify the analysis of output photon number distributions of the DFT by taking a closer look at the symmetry between the input modes and its effects on the relation between input and output in the DFT.

\section{Eigenstates of the mode shift operation}
\label{sec::DFT}

The DFT in Eq. \eqref{eq::DFT_transformation} is based on a cyclic ordering of the modes given by the mode index $m$ which is a modular value running from $m=0$ to $m=M-1$. This ordering of the modes can be used to define a cyclic mode shift operation which transforms the creation operator of the $m$-th mode creation operator into the creation operator of the $(m+1)$-th mode,
\begin{equation}
\label{eq::mode_shifting_def}
  \hat{S} \hat{a}^{\dagger}_{m} \hat{S}^{\dagger} = \hat{a}^{\dagger}_{m+1},
\end{equation}
where $m$ is modular in $M$, so that $\hat{a}^\dagger_{M-1}$ is transformed into $\hat{a}^\dagger_0$. When applied directly to a photon number state, the mode shift operator simply moves the photon number distribution accordingly, e.g. $\hat{S}^{2}\ket{110}= \hat{S}\ket{011} = \ket{101}$.
One can then identify sets of cyclic patterns $\mathcal{E}_{\boldvec{p}}$, such that all of the states in the set can be transformed into each other by multiple applications of the shift operation $\hat{S}$,
\begin{equation}
\label{eq::p-pattern_class_def}
  \mathcal{E}_{\boldvec{p}} = \{ \hat{S}^{\Delta m}\ket{\boldvec{p}}
  :\Delta m=0,...,M-1
  \}.
\end{equation}
The whole set can then be identified by a single representative photon number distribution pattern $\boldvec{p}$. In the following, we will refer to the set of states $\mathcal{E}_{\boldvec{p}}$ as the input pattern class $\boldvec{p}$. Note that the cardinality of most input pattern classes is $M$, since it usually takes $M$ applications of $\hat{S}$ to return to the original pattern. However, some input pattern classes have a lower cardinality because the symmetry of the original pattern reduces the number of shifts needed to return to the original pattern, resulting in integer fractions of the original cardinality, $M/2$, $M/3$, and so on. For example, the input pattern class $\mathcal{E}_{1010}=\{\ket{1010},\ket{0101}\}$ has a cardinality of $M/2=2$, since only two shifts are needed to return the pattern $1010$ to itself.
The Hilbert subspace spanned by the photon number basis $\mathcal{E}_{\boldvec{p}}$ of the input pattern class $\boldvec{p}$ will be denoted as $\mathbb{H}_{\boldvec{p}}$ in the rest of this paper, while the cardinality of the input pattern class $\boldvec{p}$ is denoted by $d_{\boldvec{p}}$.

We can now consider the effect of a mode shift operation performed in the input of a DFT on the output photon number distributions observed after the DFT. According to the relation between input creation operators and output creation operators described in Eq.(\ref{eq::DFT_transformation}), the effect of the output mode shift operation is given by
\begin{equation}
\label{eq::phase_shift_inverse_DFT_of_mode}
  \hat{S}\left(\hat{U}_{\mathrm{F}}^{\dagger}\hat{a}_{k}^{\dagger}\hat{U}_{\mathrm{F}}\right)\hat{S}^\dagger
  =
  e^{i\frac{2\pi}{M}k} \left(\hat{U}_{\mathrm{F}}^{\dagger}\hat{a}_{k}^{\dagger}\hat{U}_{\mathrm{F}}\right).
\end{equation}
Thus, the mode shift operation performed in the input of the DFT results in a phase shift for each output mode $k$.
Specifically, the phase shift is given by $2 \pi k/M$, which is proportional to the mode index $k$.
The eigenstates of a phase shift in a given mode are the photon number states of that mode.
In the present case, the output photon number states $\hat{U}_{\mathrm{F}}^{\dagger}\ket{\boldvec{n}}$ are the eigenstates of the mode shift operator $\hat{S}$, where the eigenvalues are given by the products of the phase factors $\exp(i2 \pi n_{k} k/M)$ of the $M$ output modes.
The eigenvalue equation then reads
\begin{equation}
\label{eq::mode_shifting_of_inverse_DFT}
  \hat{S} \; \hat{U}_{\mathrm{F}}^\dagger \ket{\boldvec{n}} = e^{i\frac{2\pi}{M}K(\boldvec{n})} \;\hat{U}^{\dagger}_{\mathrm{F}}\ket{\boldvec{n}},
\end{equation}
where $K(\boldvec{n})$ is the modular value of the total mode index of the photon number state $\ket{\boldvec{n}}$ given by
\begin{equation}
\label{eq::K-value_def}
  K(\boldvec{n})=\sum_{k=0,...,M-1}n_{k}k \pmod M.
\end{equation}
A value of $K=0$ indicates that the pattern is centered around the $k=0$ mode, so that for any photon shifted to a higher $k$ value, another photon is shifted to a lower $k$ value. A non-zero $K$ value can be obtained by shifting one of the photons in a $K=0$ pattern to the next higher mode. The $K$ value is therefore a measure of the total unbalanced mode shifts in the distribution of photons over the output modes.
For example, the patterns $0101$, $2000$ and $0020$ are symmetrically distributed around the $0$-th mode and have a total value of $K=0$, while the patterns $0011$, $1100$ can be obtained by shifting one photon in the $K=0$ patterns to the next mode and therefore have a total value of $K=1$.

As shown by Eq. \eqref{eq::mode_shifting_of_inverse_DFT}, the statistics of the measurement outcomes detected after the DFT are not changed by a mode shift operation on the input,
\begin{equation}
\label{eq::conservation_under_mode_shifting}
  |\braket{\boldvec{n}|\hat{U}_{\mathrm{F}} \hat{S}^{\Delta m}|\psi }|^2
  =
  |\braket{\boldvec{n}|\hat{U}_{\mathrm{F}}|\psi }|^2,
\end{equation}
where $\Delta m$ is a completely arbitrary integer. This relation has direct implications for the cyclic pattern classes $\mathcal{E}_{\boldvec{p}}$, since the elements in these classes are related to each other by input mode shifts. We can therefore conclude that each state in a pattern class produces the same output statistics after the DFT is applied to it, which means that photon detection after the DFT cannot distinguish between different input states from the same pattern class.
To identify the states within the Hilbert space of an input pattern class that are distinguished by the output measurement, we have to consider the eigenstates of the mode shift operators for that specific pattern class,
\begin{equation}
\label{eq::S-eigenstates_def}
  \hat{S} \ket{\mathcal{E}_{\boldvec{p}},K} = e^{i\frac{2\pi}{M} K} \ket{\mathcal{E}_{\boldvec{p}},K}.
\end{equation}

Since the application of a mode shift operation on each element of the pattern class produces the next element in the class, the eigenstates are equal superpositions of all the elements in the class with the phases determined by $K$,
\begin{equation}
\label{eq::Kp-basis_def}
  \ket{\mathcal{E}_{\boldvec{p}},K} :=
  \frac{1}{\sqrt{d_{\boldvec{p}}}}\sum_{\Delta m=0}^{d_{\boldvec{p}}-1} e^{-i\frac{2\pi}{M}K \Delta m} \hat{S}^{\Delta m} \ket{\boldvec{p}},
\end{equation}
where $d_{\boldvec{p}}$ is the cardinality of the pattern class $\boldvec{p}$. Thus, in the Hilbert space $\mathbb{H}_{\boldvec{p}}$ of the pattern class $\boldvec{p}$, the basis $\{\ket{\mathcal{E}_{\boldvec{p}},K}\}$ and the photon number basis $\{\hat{S}^{\Delta m}\ket{\boldvec{p}}\}$ are mutually unbiased.
The possible $K$ values within the pattern class depend on the cardinality $d_{\boldvec{p}}$ and are multiples of $M/d_{\boldvec{p}}$,
\begin{equation}
\label{eq::cardinality_and_K}
  K = 0, \frac{M}{d_{\boldvec{p}}}, 2\frac{M}{d_{\boldvec{p}}}, ..., M-\frac{M}{d_{\boldvec{p}}}.
\end{equation}
For example,  the set of eigenstates of the $\hat{S}$ operator in the input pattern class $1010$ has a cardinality of $2$, and $K$-values of $0$ and $2$. Specifically, the eigenstates $\ket{\mathcal{E}_{\boldvec{p}},K}$ are given by
\begin{align}
\label{eq::S-eigenstates_1010-class}
  \ket{\mathcal{E}_{1010},0} = 
  \frac{1}{\sqrt{2}}(\ket{1010}+\ket{0101}), \;\;\; 
  \ket{\mathcal{E}_{1010},2} = 
   \frac{1}{\sqrt{2}}(\ket{1010}-\ket{0101}).
\end{align}

Eq. \eqref{eq::Kp-basis_def} clarifies the relation between $K$-values of the output photon number statistics and the quantum coherence between the different photon number distributions $\hat{S}^{\Delta m}\ket{\boldvec{p}}$ in the input states of the DFT. Each output photon number state $\ket{\boldvec{n}}$ has its own specific $K$-value, and therefore exists in the eigenspace of the mode shift operator $\hat{S}$ spanned by the corresponding $\hat{S}$-eigenstates $\ket{\mathcal{E}_{\boldvec{p}},K}$ of the different input pattern classes.
It is therefore possible to express the measurement outcomes $\{\hat{U}_{\mathrm{F}}\ket{\boldvec{n}}\}$ as superpositions of pattern class states $\ket{\mathcal{E}_{\boldvec{p}},K}$ with the same $K$-value ,
\begin{equation}
\label{eq::inverse_DFT_of_photon_number_state}
  \hat{U}_{\mathrm{F}}^{\dagger}\ket{\boldvec{n}} = \sum_{\boldvec{p}:|\boldvec{p}|=|\boldvec{n}|} c_{(\boldvec{n},\boldvec{p})}\ket{\mathcal{E}_{\boldvec{p}},K(\boldvec{n})},
\end{equation}
where the sum runs over all pattern classes that have the same total photon number as $\boldvec{n}$ and include a state with a $K$-value of $K(\boldvec{n})$ according to Eq. \eqref{eq::cardinality_and_K}.
For example, $K=1$ only occurs in pattern classes with the full cardinality of $M$.
For a $2$-photon $4$-mode state, this means that the sum runs only over the pattern classes $\mathcal{E}_{2000}$ and $\mathcal{E}_{1100}$.
Specifically, the output states are given by
\begin{align}
\label{eq::eg_output_st_in_input}
  \hat{U}_{\mathrm{F}}^{\dagger}\ket{1100} & =
  \frac{1}{\sqrt{2}}\left(
  \ket{\mathcal{E}_{2000},1} + e^{-i\frac{\pi}{4}}\ket{\mathcal{E}_{1100},1}\right),
  \nonumber \\
  \hat{U}_{\mathrm{F}}^{\dagger}\ket{0011} & =
  \frac{1}{\sqrt{2}}\left(
  \ket{\mathcal{E}_{2000},1} - e^{-i\frac{\pi}{4}} \ket{\mathcal{E}_{1100},1}\right).
\end{align}
Experimentally, the observation of output photon number distribution $\ket{\boldvec{n}}$ distinguishes between different $K$-values, which in turn correspond to different superpositions of input photon number distributions, as shown by Eq. \eqref{eq::Kp-basis_def}.
The observation of a specific $K$-value therefore has direct implications for the quantum coherence between different input photon number states.

We summarize these results by formulating the \emph{mode shift rule of DFTs}, which states that the $K$-values of photon number distributions $\boldvec{n}$ obtained in the output of a DFT always distinguish the different eigenspaces of a mode shift operation in the input.
The mode shift rule can be used to identify the effect of quantum coherences between photon number states in the input on the photon number statistics observed in the output.
If a single $K$-value is observed in the output, it is possible to
associate this $K$-value with a specific phase relation between input photon number distributions related to each other by a mode shift operation, as shown in Eq. \eqref{eq::Kp-basis_def}.
Although it is not possible to identify the pattern classes from which the $K$-value originates, the observation of a specific $K$-value is always associated with off-diagonal elements of the density matrix expressed in the input photon number basis, and the probabilities of different $K$-values are determined by summations of the off diagonal elements representing coherent superpositions of input photon number states.

It may be worth noting that there is one special case where the mode shift rule relates input photon numbers directly to output photon numbers. This occurs in pattern classes of cardinality $1$, which have the same photon number in each input mode.
In this case, the mode shift rule reproduces the suppression laws of DFTs introduced in \cite{TichyEtAlBuchleitner2010-MultiBS,TichyEtAlBuchleitner2012-MPInteference},
which predict zero probability for all output patterns that do not have a $K$-value of $0$.
However, this is the only case in which the mode shift rule does not involve quantum coherent superpositions of different input photon number distributions.

For the evaluation of multi-mode entanglement by the correlations of outputs after local DFTs, the general mode shift rule is needed to evaluate the quantum coherence between photon number components.
Due to the entanglement, the quantum coherences observed in the two systems will be correlated, and these quantum correlations are observable as correlations between the $K$-values observed in the output of the DFTs. In the following, we will apply the mode shift rule to derive the $K$-value correlations and use the results to derive entanglement witnesses based on photon detection before and after local DFTs.

\section{Entanglement criterion}
\label{sec::ent_crit}

The merit of applying the mode shift rule to the analysis of the photon number statistics observed after a DFT is that the large amount of information in the photon number distribution $\boldvec{n}$ obtained in the output measurement can be reduced to a single $K$-value using Eq.(\ref{eq::K-value_def}). In the case of the entangled state $\ket{\phi_{N,M-N}}$ between two multi-mode systems of different local photon numbers $N$ and $M-N$, this makes it possible to efficiently analyze correlations between a wide variety of very different photon number distributions observed at $A$ and at $B$.

The application of the DFT to characterize the entangled state $\ket{\phi_{N,M-N}}$ is motivated by the fact that the beam splitting procedure is completely symmetric in the modes $M$, so that a rearrangement of mode labels has no effect whatsoever on the entangled state, as long as the lables in system $A$ correspond to the lables in system $B$. It is therefore obvious that a simultaneous mode shift operation performed in $A$ and in $B$ does not change the state at all, making $\ket{\phi_{N,M-N}}$ an eigenstate of this operation with an eigenvalue of one,
\begin{equation}
\label{eq::K-sum}
  \hat{S}^{(A)}\otimes\hat{S}^{(B)}\ket{\phi_{N,M-N}} = \ket{\phi_{N,M-N}}.
\end{equation}
If the entangled state $\ket{\phi_{N,M-N}}$ is represented in the eigenstate basis of the local mode shift operations $\hat{S}^{(A)}\otimes\hat{S}^{(B)}$, Eq. \eqref{eq::K-sum} requires that only basis states with a modular sum of $K_{A}+K_{B}=0$ contribute to the entangled state.
Within each pattern class, the perfect correlation between input photon number patterns therefore transforms into a perfect correlation between $K$-values, so that the entangled state for total photon numbers of $N$ and $M-N$ can be written as
\begin{align}
\label{eq::ent_st_NANB_in_kp}
  \ket{\phi_{N,M-N}}
  = \frac{1}{\sqrt{\binom{M}{N}}} \sum_{\boldvec{p}\in \mathcal{P}_{\phi}^{(A)}}
  \sum_{K}
  \ket{\mathcal{E}_{\boldvec{p}},K}\ket{\mathcal{E}_{\bar{\boldvec{p}}},-K},
\end{align}
where $\mathcal{P}_{\phi}^{(A)}$ denotes the set of pattern classes of $N$ photons with zero or one photon in each mode and $\bar{\boldvec{p}}$ is the complementary pattern class, obtained by exchanging zero photon with one photon and vice versa.
It should be noted that the initial patterns $\boldvec{p}$ and $\bar{\boldvec{p}}$ are different from each other, resulting in different global phases for $\ket{\mathcal{E}_{\boldvec{p}},K}$ and $\ket{\mathcal{E}_{\bar{\boldvec{p}}},K}$ according to Eq. \eqref{eq::Kp-basis_def}, when $\boldvec{p}$ and $\bar{\boldvec{p}}$ belong to the same patten class.
For example, for $2$-photons in $4$-modes $\ket{\mathcal{E}_{1100},1} = - \ket{\mathcal{E}_{0011},1}$.

Within each pattern class subspace, the eigenstates of the mode shift operators and the photon number states are mutually unbiased.
This means that separable states can never have both opposite $K$-values $(K,-K)$ in the local output modes and complementary photon number patterns $(\boldvec{n},\bar{\boldvec{n}})$ in the local input modes.
For mixed states, this puts a well defined limit on the possible simultaneous correlations in the $K$-values and in the input photon number distributions.
The observation of both correlations $(K,-K)$ between the $K$-values of the output photon number distribution of the DFT and the complementary correlations $(\boldvec{n},\bar{\boldvec{n}})$ in the original photon number basis is therefore an experimentally accessible signature of the entanglement between $A$ and $B$.
It is thus possible to verify the entanglement of the state $\ket{\phi_{N,M-N}}$ by quantifying the experimentally observed correlations between the photon number patterns in the original modes, together with the correlations between $K$-values observed after the DFT has been performed on both multi-mode systems.
If these correlations exceed the bound satisfied by all separable states, the experimental data confirms the successful generation of entanglement between the multi photon systems at $A$ and at $B$.

The correlations in the photon detection in the input modes can be quantified by the fidelity $F_{\boldvec{n}}$, which is given by the probability of observing the expected correlation $(\boldvec{n},\bar{\boldvec{n}})$ between the input modes at $A$ and at $B$. By summing over the probability of all correctly correlated combinations of photon number distributions, we obtain
\begin{equation}
\label{eq::E_corr_prob}
  F_{\boldvec{n}} = \sum_{n_{m}\leq1} \braket{\boldvec{n}\bar{\boldvec{n}}|\hat{\rho}_{\text{exp.}}|\boldvec{n}\bar{\boldvec{n}}},
\end{equation}
where $\hat{\rho}_{\text{exp.}}$ is the density operator of the experimentally generated state.
Similarly, the correlations between the $K$-values of the photon number distributions detected after DFTs have been performed in both systems can be quantified by the fidelity $F_{K}$, which is given by the probability of observing the correct correlation $(K,-K)$ between the output modes at $A$ and at $B$. By summing over the probability of all correctly correlated combinations of photon number distributions, we obtain
\begin{align}
\label{eq::K_corr_prob}
  F_{K} =
  \sum_{K(\boldvec{n}_{A})+K(\boldvec{n}_{B})=0}
  \braket{\boldvec{n}_{A}\boldvec{n}_{B}|\hat{U}_{\mathrm{F}}\otimes\hat{U}_{\mathrm{F}}\hat{\rho}_{\text{exp.}} \hat{U}_{\mathrm{F}}^{\dagger}\otimes\hat{U}_{\mathrm{F}}^{\dagger}|\boldvec{n}_{A}\boldvec{n}_{B}},
\end{align}
where $K_A$, $K_B$ are the $K$-values associated with the photon number distributions $\boldvec{n}_{A}$, $\boldvec{n}_{B}$ respectively.

For the ideal entangled state $\ket{\phi_{N,M-N}}$, these fidelities are both equal to $1$, achieving the maximal value of $2$ for the sum of both fidelities. To obtain a quantitative criterion for experimental entanglement detection, we have to identify the upper bound of the fidelity sum for separable states. To do so, it is helpful to express the fidelities as expectation values of operators,
\begin{equation}
  F_{\boldvec{n}} + F_{K} = \tr(\hat{C}_{\boldvec{n}}\hat{\rho}_{\text{exp.}}) + \tr(\hat{C}_{K}\hat{\rho}_{\text{exp.}}).
\end{equation}
These operators are projectors on the Hilbert spaces associated with the expected correlations between the measurement outcomes. For the input photon number correlation, this projection operator can be given by
\begin{equation}
  \hat{C}_{\boldvec{n}} = \sum_{n_{m}\le1}\projector{\boldvec{n}\bar{\boldvec{n}}},
\end{equation}
where each local state is associated with a specific pattern class. It is therefore possible to re-organize the sum into pattern classes, so that the correlations between pattern classes $\boldvec{p}$ and complementary pattern classes $\bar{\boldvec{p}}$ are separated from the correlations between pattern shifts $m$,
\begin{equation}
\label{eq::Cn_in_patterns}
  \hat{C}_{\boldvec{n}} = \sum_{\boldvec{p}} \sum_{\Delta m=0}^{d_{\boldvec{p}}-1} (\hat{S}\otimes\hat{S})^{\Delta m} \projector{\boldvec{p}\bar{\boldvec{p}}} (\hat{S}^{\dagger}\otimes\hat{S}^{\dagger})^{\Delta m}.
\end{equation}
Importantly, the correlation is always zero if the photon number distributions observed in the output belong to different pattern classes.
Oppositely, the fidelity of the correlation between $K$-values in the output modes of the DFT is completely independent of the pattern class combination. If the sum over pattern classes is separated from the sum over the possible combinations of $K$ and $-K$, the operator for the output photon number correlation reads
\begin{align}
\label{eq::Ck_in_patterns}
  \hat{C}_{K} = 
  \sum_{\boldvec{p}_A, \boldvec{p}_B} \sum_{K} \projector{\mathcal{E}_{\boldvec{p}_A},K;\mathcal{E}_{\boldvec{p}_B},-K}.
\end{align}
Since this operator commutes with projections on specific pattern class subspaces, we can evaluate the correlations between the $K$-values separately for each combination of pattern classes $\boldvec{p}_{A}$ and $\boldvec{p}_{B}$.
To find an experimental criterion for the detection of entanglement, we now consider the maximal fidelity sum that can be obtained
if the imperfect state generated in the experiment is separable and contains no entanglement whatsoever,
\begin{align}
\label{eq::corr_fidelity_upper_bound_0}
   \max_{\text{sep.}} (F_{\boldvec{n}}+ F_{K})
   = \max_{\rho_{sep.}}(\tr((\hat{C}_{\boldvec{n}}+\hat{C}_{K})\hat{\rho}_{\text{sep.}})) =
   B_{\text{sep.}}.
\end{align}
If the sum of the correlation fidelities exceeds this boundary, the state $\hat{\rho}_{\text{exp.}}$ generated in the experiment is definitely entangled.

Since the operators representing the correlations between the input photon number distributions $\boldvec{n}$ and the correlations between the $K$-values of the output photon number distributions both commute with projectors onto the pattern class subspaces, we can express the fidelities as a product of pattern class probability $P(\boldvec{p}_{A},\boldvec{p}_{B})$ and the fidelity for the projection of the state into that pattern class combination. Here, the pattern class probability is given by
\begin{equation}
P(\boldvec{p}_{A},\boldvec{p}_{B}) = \tr(\hat{\Pi}_{\boldvec{p}_A,\boldvec{p}_{B}} \hat{\rho}_{\mathrm{exp.}})
\end{equation}
and the conditional pattern class states are given by
 \begin{equation}
\hat{R}_{\boldvec{p}_A,\boldvec{p}_{B}} = \frac{\hat{\Pi}_{\boldvec{p}_A,\boldvec{p}_{B}} \hat{\rho}_{\mathrm{exp.}}\hat{\Pi}_{\boldvec{p}_A,\boldvec{p}_{B}}}{P(\boldvec{p}_{A},\boldvec{p}_{B})},
\end{equation}
where $\hat{\Pi}_{\boldvec{p}_{A}, \boldvec{p}_{B}}$ is the projection operator on the $(\boldvec{p}_{A}, \boldvec{p}_{B})$-pattern subspace $\mathbb{H}_{\boldvec{p}_{A}}\otimes\mathbb{H}_{\boldvec{p}_{B}}$.
We can then express the Fidelities $F_{\boldvec{n}}$ and $F_K$ as an average over the fidelities in the different pattern classes, as given by
\begin{align}
\label{eq::ent_witness_diagonalized_in_p}
  F_{\boldvec{n}} + F_K 
  =  \sum_{\boldvec{p}_{A},\boldvec{p}_{B}} P(\boldvec{p}_{A},\boldvec{p}_{B})
  \tr((\hat{C}_{\boldvec{n}} + \hat{C}_K)\hat{R}_{\boldvec{p}_A,\boldvec{p}_{B}}).
\end{align}
Since the maximum of a statistical average cannot be larger than the maximal individual contribution, we can identify all relevant bounds on $F_{\boldvec{n}} + F_K$ by selecting the pattern class subspace within which the maximal fidelity sum is achieved. For that purpose, we need to evaluate the specific limits of the fidelity sum in each pattern class subspace.

A particularly simple limit is obtained when the pattern classes in system A and system B are not complementary ($\boldvec{p}_{A}\neq\bar{\boldvec{p}}_{B}$), so that according to Eq. \eqref{eq::E_corr_prob} $F_{\boldvec{n}}$ is zero.
Since $F_K \le 1$, the sum of the fidelities cannot exceed one for any state, entangled or separable.
The maximal value of $F_{\boldvec{n}} + F_K =2$ associated with the ideal entangled state can only be achieved in complementary pattern classes.
For separable states $\hat{R}_{\boldvec{p},\bar{\boldvec{p}}}{(\mbox{sep.})}$ in complementary pattern subspaces ($\boldvec{p}_{A}=\bar{\boldvec{p}}_{B}$), the photon number basis $\{\hat{S}^{\Delta m} \ket{\boldvec{p}}\}$ and the $K$-basis $\{\ket{\mathcal{E}_{\boldvec{p}},K}\}$ are mutually unbiased.
This means the eigenstates of $\hat{C}_{\boldvec{n}}$ with eigenvalues of $1$ are equally distributed over all possible combinations of $K_{A}$ and $K_{B}$, so that the probability of opposite $K$-values $K_{A}=-K_{B}$ is $1/d_{\boldvec{p}}$.
Likewise, eigenstates of $\hat{C}_{K}$ with eigenvalues of $1$ have a probability of $1/d_{\boldvec{p}}$ for complementary photon number distributions $(\boldvec{n},\bar{\boldvec{n}})$.
It can be shown that no other product states exceed the sum of fidelity $F_{\boldvec{n}} + F_{K}$ achieved by these eigenstates \cite{WuYuMolmer2009-EntUnRelMUB, SpenglerHuberEtAlHiesmayr2012-EntWitViaMUB}.
Therefore the bound for separable states  $\hat{R}_{\boldvec{p},\bar{\boldvec{p}}}{(\mbox{sep.})}$ in the complementary pattern subspaces $(\boldvec{p},\bar{\boldvec{p}})$ is given by
\begin{equation}
\label{eq::ent_witness_in_comp_p_class}
  \tr\left((\hat{C}_{\boldvec{n}} + \hat{C}_K)\hat{R}_{\boldvec{p},\bar{\boldvec{p}}}{(\mbox{sep.})}\right)
  \le
  1 + \frac{1}{d_{\boldvec{p}}},
\end{equation}
where $d_{\boldvec{p}}$ is the cardinality of the $\boldvec{p}$-pattern class.
We can summarize the results for complementary and non-complementary pattern classes in a single inequality as follows,
\begin{equation}
\label{eq::ent_witness_in_p_class}
  \tr\left((\hat{C}_{\boldvec{n}} + \hat{C}_K)\hat{R}_{\boldvec{p}_{A},\boldvec{p}_{B}}{(\mbox{sep.})}\right)
  \le
  1 + \frac{1}{d_{\boldvec{p}_{A}}}\delta_{\boldvec{p}_{A},\bar{\boldvec{p}}_{B}}.
\end{equation}

According to Eq. \eqref{eq::ent_witness_diagonalized_in_p}, the fidelities for arbitrary states are obtained by averaging over the fidelities of the contributions from different pattern spaces. Thus the highest possible value of $F_{\boldvec{n}} + F_{K}$ is achieved by a separable state from the subspace of complementary pattern classes $(\boldvec{p},\bar{\boldvec{p}})$ with the lowest possible cardinality $d_{\boldvec{p}}$. As a result, the upper bound of the fidelity sum for all separable states is given by
\begin{equation}
\label{eq::corr_fidelity_upper_bound}
  \max_{\text{sep.}} (F_{\boldvec{n}} + F_{K})
  =
  1+\frac{1}{ \min_{\boldvec{p}}d_{\boldvec{p}} }.
\end{equation}
As mentioned above, any state that exceeds this bound is necessarily entangled. We can therefore verify the bipartite entanglement between the two multi-mode systems at $A$ and at $B$ by evaluating the correlations between photon distributions detected in the input modes and the correlations between the $K$-values of the photon distributions detected in the output modes of a DFT. If the experimentally determined values exceed the bound given by Eq.\eqref{eq::corr_fidelity_upper_bound} the generation of entanglement between the two multi-mode systems has been verified.

The upper bound derived above makes no use of any information about the actual pattern class distribution of the generated state. This may be a significant disadvantage when most pattern classes of the state have a much higher cardinality than the minimal cardinality used for the bound. It may therefore be useful to consider the actual distribution of cardinalities of pattern classes for a specific number of modes $M$. Specifically, a cardinality of $d_{\boldvec{p}}=M$ means that the photon number distribution $\boldvec{p}$ is different for each mode shift, while a cardinality of $d_{\boldvec{p}}=M/\nu$ means that the photon number distribution $\boldvec{p}$ returns to the original distribution after only $M/\nu$ mode shifts because the same pattern is repeated $\nu$ times along the $M$ modes. Existence of pattern classes with lower cardinalities thus requires that $M$ can be factorized into $\nu$ and $M/\nu$. If the total photon number $M$ is a prime number, then every local pattern class has the cardinality $M$ and the upper bound in Eq. \eqref{eq::corr_fidelity_upper_bound} is optimal for every local pattern class $(\boldvec{p}_{A},\boldvec{p}_{B})$. However, prime numbers are less and less common as the number of modes increases, so that a limitation to prime numbers may be inconvenient. In addition, an even number of modes (or even a power of two for the number of modes) may be convenient for the design of the experimental setup. We should therefore also consider the worst case scenario, which occurs when an even total photon number $M$ is split into equal photon numbers of $(M/2,M/2)$. The lowest cardinality of a pattern class of $M/2$ photons in $M$ modes is the cardinality of $2$ of the pattern class $\mathcal{E}_{1010\cdots10}$ which is complementary to itself. In this case, the bound derived above is definitely far from optimal, resulting in a maximal value of $3/2$ that is only achieved by separable states from the two dimensional subspaces of a single pattern class. It is easy to see from the correlation measurements of the input photon number distributions that only a very small fraction of the photons is found in this very small pattern class. We should therefore consider an improvement of the bound that includes the actual pattern class statistics observed in the experiment.

\section{Tighter bounds based on pattern class statistics}
\label{sec::tighter_bound}

As mentioned above, we actually obtain detailed information about the pattern class distribution in the photon detection measurement of the input modes of the local DFTs. As a consequence, we can determine the probability distribution $P(\boldvec{p}_{A},\boldvec{p}_{B})$ from the experimental data.
Since the bounds for the different combinations of pattern classes given by Eq. \eqref{eq::ent_witness_in_p_class} must be satisfied separately by each combination $(\boldvec{p}_{A},\boldvec{p}_{B})$, the bound for a specific probability distribution $P(\boldvec{p}_{A},\boldvec{p}_{B})$ is achieved by a mixture of separable states from each pattern class that achieve the bound for that pattern class. This bound is therefore simply equal to the average of the bounds for the individual pattern class combinations,
\begin{equation}
\label{eq::ent_witness_optimized_derive_1}
  \max_{\text{sep.}} (F_{\boldvec{n}} + F_{K})
  =
  1+\sum_{\boldvec{p}}\frac{1}{d_{\boldvec{p}}} P(\boldvec{p},\bar{\boldvec{p}}).
\end{equation}
Note that the bound is one if the pattern classes are not complementary, so the sum in Eq. \eqref{eq::ent_witness_optimized_derive_1} run only over the complementary pattern class combinations.

Usually, bounds are formulated as maximal or minimal values achieved by the experimental results under specific conditions. It is therefore useful to consider the measurement-dependent term on the right-hand side of Eq. \eqref{eq::ent_witness_optimized_derive_1} on an equal footing to the fidelities. For this purpose, we define the statistical average of $1/d_{\boldvec{p}}$ obtained from the experimentally observed distribution of pattern classes as the the pattern class defect $D_{\boldvec{p}}$ with
\begin{equation}
\label{eq::def_Dp}
  D_{\boldvec{p}} = \sum_{\boldvec{p}}\frac{1}{d_{\boldvec{p}}}P(\boldvec{p},\bar{\boldvec{p}}).
\end{equation}
By including the pattern class defect in the experimental evaluation of the quantum state, we can formulate a new bound on separable states that reads
\begin{equation}
\label{eq::ent_criterion_optimized}
  \max_{\text{sep.}} (F_{\boldvec{n}} - D_{\boldvec{p}} + F_{K})  = 1,
\end{equation}
where $F_{\boldvec{n}}$ and $D_{\boldvec{p}}$ are obtained from the measurement on the input modes, and $F_{K}$ is obtained from the measurements in the output of local DFTs.
It is also possible to express the pattern class defect $D_{\boldvec{p}}$ as the expectation value of an operator,
\begin{equation}
  D_{\boldvec{p}} = \tr(\rho_{\text{exp.}}\hat{C}_{\boldvec{p}}) := \tr\left(\rho_{\text{exp.}}\sum_{\boldvec{p}:n_{m}\le1}\frac{1}{d_{\boldvec{p}}} \hat{\Pi}_{\boldvec{p},\bar{\boldvec{p}}}\right),
\end{equation}
where the sum runs over all the local pattern classes $\boldvec{p}$ that have corresponding complementary pattern class, i.e. the photon number in each mode does not exceeds 1, $n_{m}\le1$.
For completeness, we can also express the entanglement criterion by a witness operator $\hat{W}$, defined so that any positive expectation value of $\hat{W}$ indicates entanglement. In terms of the operators defined previously, this entanglement witness is given by
\begin{equation}
\hat{W} = \hat{C}_{\boldvec{n}} - \hat{C}_{\boldvec{p}} + \hat{C}_K - \id.
\end{equation}
The maximal eigenvalues of this entanglement witness are associated with the pattern classes of maximal cardinality, $d_{\boldvec{p}}=M$. Within each pattern class of maximal cardinality $M$, there exists a maximally entangled eigenstate of $\hat{W}$ with the maximal eigenvalue of $1-1/M$.

Since the intended entangled state $\ket{\phi_{N,M-N}}$ includes pattern classes with different cardinality, it is not an eigenstate of W and has an expectation value of W that is somewhat lower than the maximal eigenvalue of $1-1/M$. Specifically, the expectation value is given by
\begin{align}
  \braket{\phi_{N,M-N}|\hat{W}|\phi_{N,M-N}} 
  =
  1 - \sum_{\boldvec{p}:n_{m}\le1} \frac{1}{d_{\boldvec{p}}} \braket{\phi_{N,M-N}|\hat{\Pi}_{\boldvec{p},\bar{\boldvec{p}}}|\phi_{N,M-N}},
\end{align}
where the probability of each pattern class with $n_m \le 1$ is proportional to its cardinality $d_{\boldvec{p}}$, so that each pattern class reduces the expectation value of the witness by the same amount,
\begin{equation}
 \frac{1}{d_{\boldvec{p}}} \braket{\phi_{N,M-N}|\hat{\Pi}_{\boldvec{p},\bar{\boldvec{p}}}|\phi_{N,M-N}} = \frac{1}{\binom{M}{N}}.
\end{equation}
Effectively, the average value of $1/d_{\boldvec{p}}$ is given by the ratio of the number of local pattern classes with $n_{m}\le1$ and the total number of local photon distributions with $n_{m}\le1$. As photon number increases, this average $D_{\boldvec{p}}$ is expected to converge on $1/M$, since the vast majority of pattern classes has a cardinality of $M$.

\section{Entanglement between two systems with two photons in four modes}
\label{sec::examples}

\def\picwidth{0.48\textwidth}
\begin{figure}[t]
  \centering
  \subfloat[][]{\includegraphics[width=\picwidth]{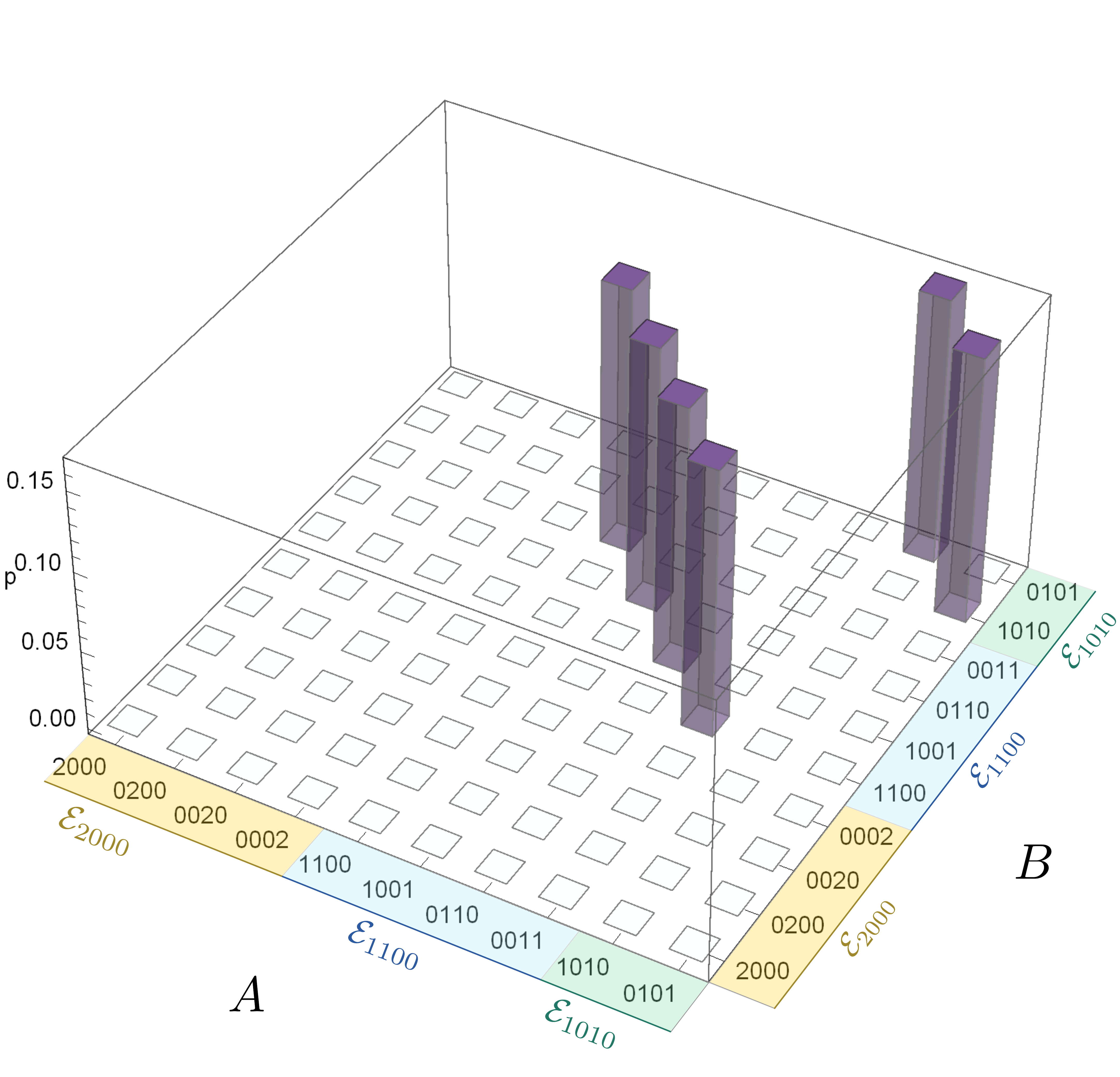}}
  \subfloat[][]{\includegraphics[width=\picwidth]{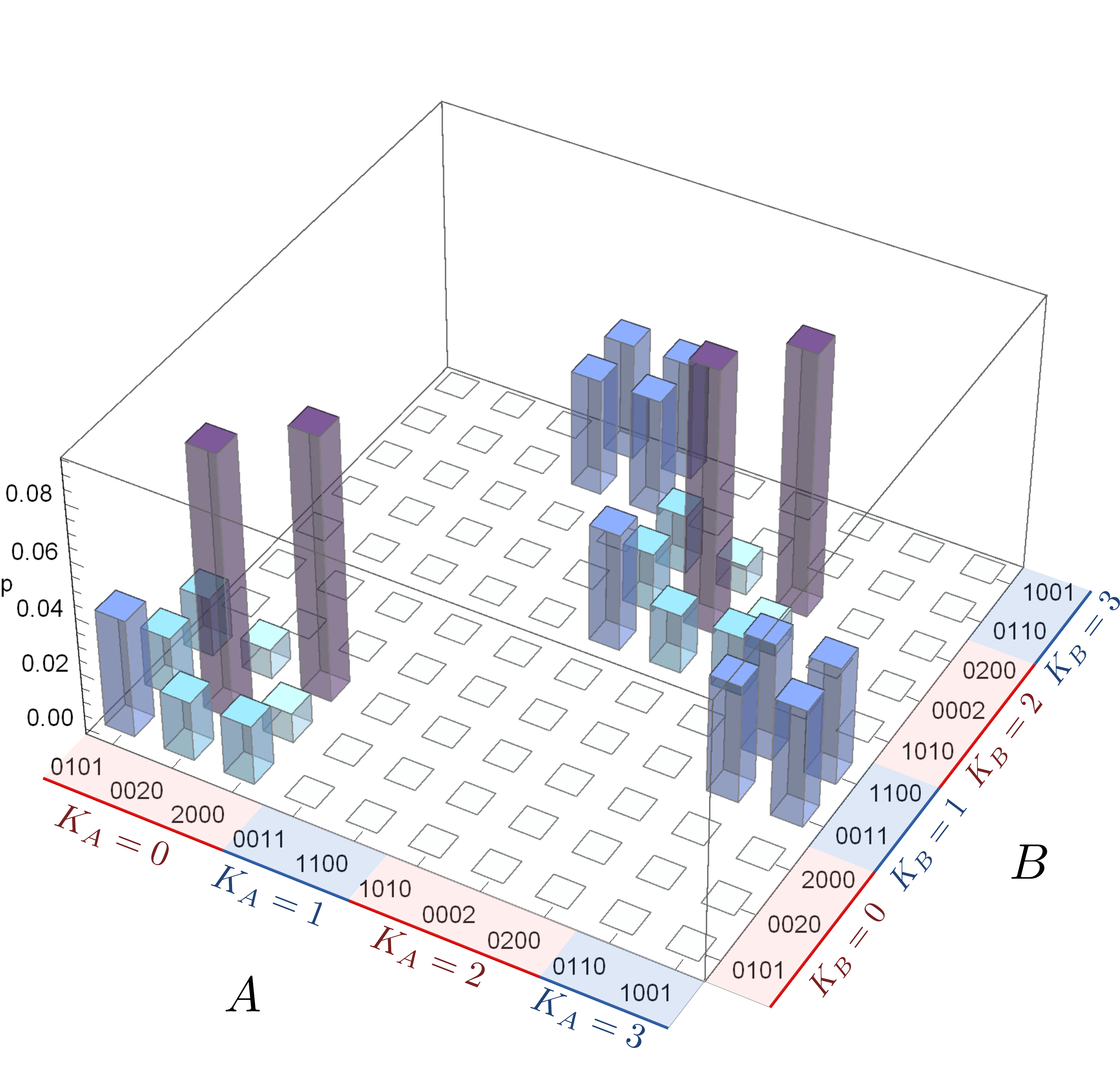}}
  \caption{\colorfig
    The probability distribution of photon number detection events for the state $\ket{\phi_{2_{A},2_{B}}}$ (a) in the original input modes and (b) in the output modes of local DFTs. The axes labled $A$ and $B$ show the photon distributions detected in each local system. In (a), the photon distributions have been arranged according to the maximal number of photons in one mode (Eq. \eqref{eq::pattern_classes_2N4M}). In (b), the distributions have been organized according to the $K$-value (Eq. \eqref{eq::K-blocks}). The first three outcomes correspond to $K=0$, the next two outcomes have $K=1$, followed by another block of three outcomes with $K=2$ and finally two outcomes with $K=3$. The outcomes are clearly grouped in correlated blocks corresponding to the expected correlations in $K$.}
  \label{fig::phi_2A2B_corr}%
\end{figure}

In this section, we will illustrate the principles explained above using the most simple non-trivial case of $M=4$ single photon inputs, where two photons are distributed to $A$ and two photons are distributed to $B$. Before the post-selection of $N=2$, the state is given by a superposition of the different photon number partitions in Eq. \eqref{eq::psi_4M}.
Since we are using linear optics and photon detection, we will always distinguish the different photon number partitions from each other. The probability of obtaining a completely useless output with zero photons in either $A$ or $B$ is $1/8$. The probability of getting a $4 \times 4$ entangled state of one photon entangled with three photons is $1/2$. In this case, there is a single pattern class in the system with one photon, and the information contributed by the multi-photon patterns in the other system is generally redundant. Thus the most interesting output is the distribution of two photons to each system, which occurs with a probability of $3/8$. However, it might be worth noting that the probability of generating usable entanglement from four single photon inputs is $7/8$, which means that single photon sources can be used to generate entanglement on demand with a success rate that is almost as high as the success rate of the initial single photon generation. In the light of the rather low success rates associated with most post-selected linear optics quantum circuits, this seems to be a noteworthy result.

We now focus on the output state $ \ket{\phi_{2_{A}, 2_{B}}}$, which is an entangled state with a Schmidt rank of $6$,
as seen in Eq. \eqref{eq::generated_ent_st_2A2B}.
The photon number distribution of this state in the input modes is shown in Fig. \ref{fig::phi_2A2B_corr}(a). In total, there are ten possible output distributions, corresponding to the possible distributions of two photons in four modes. In Fig. \ref{fig::phi_2A2B_corr}(a), these are arranged according to the three cyclic pattern classes $\boldvec{p}=2000,1100,1010$, which have the following elements
\begin{align}
\label{eq::pattern_classes_2N4M}
  \mathcal{E}_{2000} = & \{\ket{2000}, \ket{0200}, \ket{0020}, \ket{0002}\}, \nonumber
  \\
  \mathcal{E}_{1100} = & \{\ket{1100}, \ket{0110}, \ket{0011}, \ket{1001}\}, \nonumber
  \\
  \mathcal{E}_{1010} = & \{\ket{1010}, \ket{0101}\}.
\end{align}
The photon number outputs in $A$ and $B$ are perfectly correlated in the six complementary pattern pairs $(\boldvec{n},\bar{\boldvec{n}})$ with maximally one photon in each mode. The four patterns in $\mathcal{E}_{2000}$ with two photons in the same mode have a detection probability of zero.
If the photons are detected in the output modes of two local DFTs, all possible output patterns will be observed. The probabilities of the different output distributions are shown in Fig. \ref{fig::phi_2A2B_corr}(b). Here, the correlation is between $K$-values, so it is convenient to arrange the output distributions accordingly. The photon distributions with the same $K$-value are
\begin{align}
\label{eq::K-blocks}
  \mathcal{E}_{K=0} = & \{\ket{0101}, \ket{2000}, \ket{0020}\},
  &
  \mathcal{E}_{K=1} = & \{\ket{0011}, \ket{1100}\}, \nonumber
  \\
  \mathcal{E}_{K=2} = & \{\ket{1010}, \ket{0200}, \ket{0002}\},
  &
  \mathcal{E}_{K=3} = & \{\ket{0110}, \ket{1001}\} .
\end{align}
As can be seen in Fig. \ref{fig::phi_2A2B_corr}(b), the $K$-values are correlated so that the sum of the $K$-values is zero or four. Specifically, there are four blocks corresponding to $K$-values of $(0,0)$, $(1,3)$, $(2,2)$ and $(3,1)$. The remaining $12$ combinations all have a probability of zero.
The sum of correlation fidelities is therefore $F_{\boldvec{n}}+F_{K}=2$, while for the optimized entanglement criterion, we have $F_{\boldvec{n}}+F_{K}-D_{\boldvec{p}}=5/3$, since the statistics of pattern classes in Fig. \ref{fig::phi_2A2B_corr}(a) give a pattern defect value of $D_{\boldvec{p}} = 1/3$.

\begin{figure}[t]
\def\picwidth{0.45\textwidth}
  \centering
  \subfloat[][]{\includegraphics[width=\picwidth]{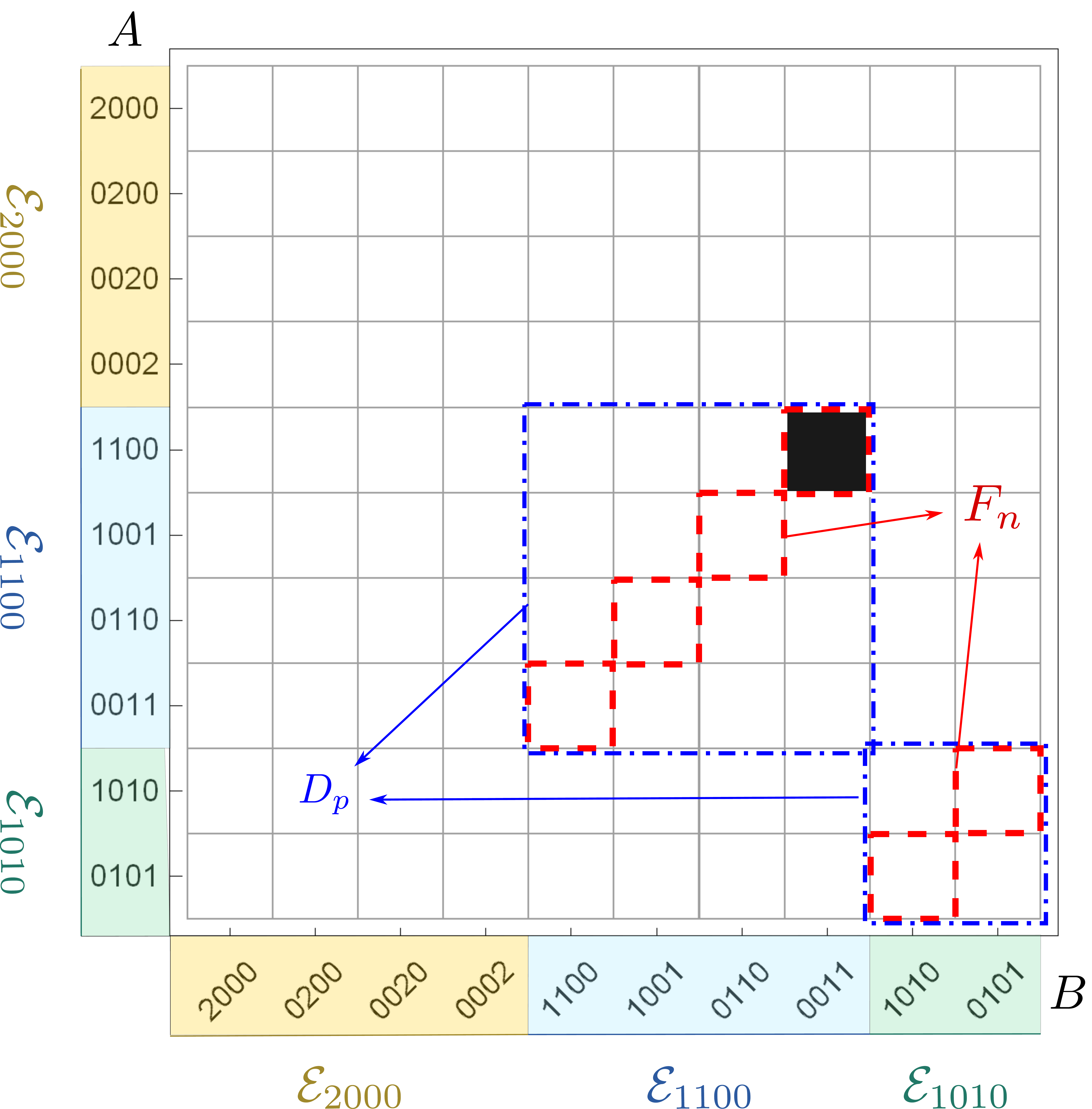}}
  \subfloat[][]{\includegraphics[width=\picwidth]{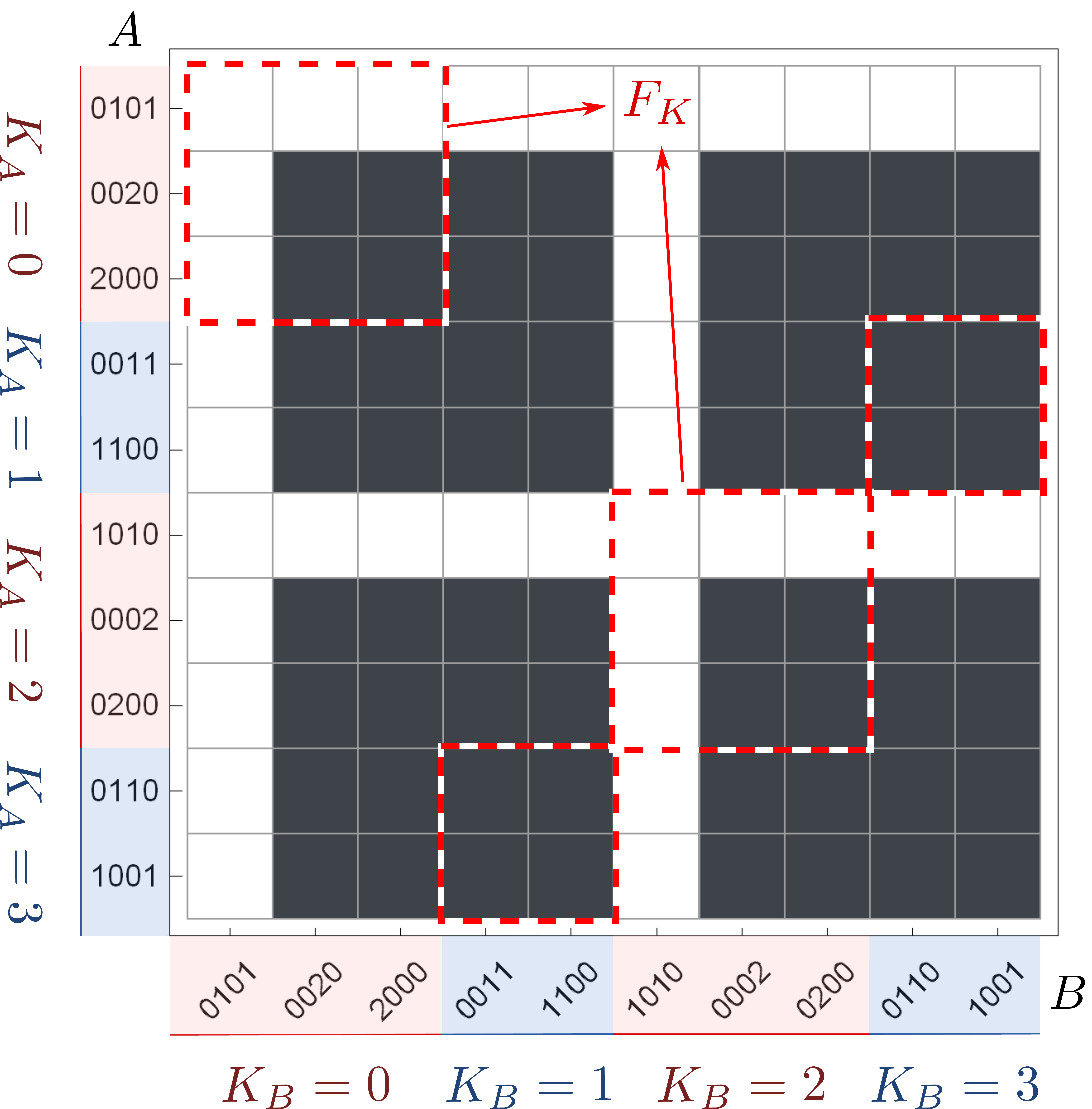}}
  \caption{\colorfig
    The statistics of photon number detection of the separable state $\ket{1100_{A},0011_{B}}$ (a) in the input modes and (b) in the output modes of two local DFTs.
  Photon distributions are arranged in the same way as in Fig. \ref{fig::phi_2A2B_corr}.
    The red dashed squares show outputs that have the correct correlation expected from the entangled state input $\ket{\phi_{2_{A},2_{B}}}$and therefore contribute to the fidelities $F_{\boldvec{n}}$ or $F_K$, respectively.
    The blue dot-dashed squares indicate the complementary pattern classes $(\boldvec{p},\bar{\boldvec{p}})$ which are used to determine the tighter bounds for separable states. In (a), the separable input state corresponds to a well-defined measurement outcome with $F_{\boldvec{n}}=1$ and $P(\mathcal{E}_{1100},\mathcal{E}_{0011})=1$. In the output modes of the DFT, the photon number distributions shown by the black squares in (b) each have a probability of $1/64$. As a result, the output mode correlation fidelity, i.e. the probability for the correlations $(K,-K)$ (red dashed squares), is $F_{K}=1/4$.
  }
  \label{fig::opt_sep_2A2B_corr_fidelity}%
\end{figure}

We can now compare the statistics of the entangled state with the statistics obtained from a separable state with $F_{\boldvec{n}}=1$. Fig. \ref{fig::opt_sep_2A2B_corr_fidelity} illustrates the statistics obtained with the product state $\ket{1100_{A},0011_{B}}$ from the pattern class $\mathcal{E}_{1100}$, which is complementary to itself.  Fig. \ref{fig::opt_sep_2A2B_corr_fidelity} (a) shows the photon number distribution associated with this state as a black square. The red dashed lines indicate the results obtained from the ideal entangled state $\ket{\phi_{2_{A},2_{B}}}$ and the blue dot-dashed squares identify the combinations of complementary pattern classes associated with the entangled state $\ket{\phi_{2_{A},2_{B}}}$. There are two pattern classes, both of which are complementary to themselves. The separable state is located in the larger pattern class, so that its value of $D_{\boldvec{p}}$ is $1/4$. For the entangled state $\ket{\phi_{2_{A},2_{B}}}$, the value is actually higher, since there are two pattern classes and six photon number combinations, resulting in a $D_{\boldvec{p}}$ value of $1/3$. Fig.  \ref{fig::opt_sep_2A2B_corr_fidelity} (b) shows the photon distribution of the separable state in the output modes of the DFT. In general, photon number states in the input modes have a completely random distribution of $K$-values. In the specific case of the $\ket{1100_{A},0011_{B}}$  state we find 64 outcomes with equal probabilities of $1/64$ each, as indicated by the black squares in Fig. \ref{fig::opt_sep_2A2B_corr_fidelity} (b). The red dashed squares indicate the results with $K_A+K_B=0$. Since the $K$-values of the local state are completely uncorrelated and equally distributed over $K=0,1,2,3$, the probability of finding the correct correlation is $F_K=1/4$. As expected, this is exactly equal to the tighter bound for separable states, $F_{\boldvec{n}} - D_{\boldvec{p}} + F_K=1$. Fig. \ref{fig::opt_sep_2A2B_corr_fidelity} thus illustrates how a separable state achieves the bound given by Eq. \eqref{eq::ent_criterion_optimized}.

The entanglement criterion applies to experimental data obtained under non-ideal circumstances. It is therefore interesting to evaluate the robustness of the state $\ket{\phi_{2_{A}, 2_{B}}}$ to potential error sources. In the absence of a specific error model, it is possible to evaluate the robustness in terms of the quantum state fidelity $F_{\text{state}}$.
Since the correlation fidelities $F_{\boldvec{n}}$ and $F_K$ for the target state are both $1$, which contributes a fraction of $F_{\text{state}}$ to the statistics, while the correlation fidelities for the errors have the minimum value of $0$, the quantum state fidelity provides a lower bound for both fidelities.
Therefore we find that
\begin{equation}
\min(F_{\boldvec{n}} + F_K) = 2 F_{\text{state}}
\end{equation}
If we apply the entanglement criterion of Eq.\eqref{eq::corr_fidelity_upper_bound} to the present case, the limit for separable states is
\begin{equation}
\min(F_{\boldvec{n}} + F_K) = 3/2.
\end{equation}
We find that this limit is always exceeded when $F_{\text{state}}>3/4$. Thus a quantum state fidelity of more than $75$ percent guarantees entanglement verification using the basic entanglement criterion of Eq.\eqref{eq::corr_fidelity_upper_bound}.
Note that the sufficiency of $75$ percent fidelity for entanglement is derived from the worst case scenario that the total mode number $M=4$ is an even number. A quantum state fidelity of $75$ percent therefore guarantees the entanglement between $A$ and $B$ for arbitrary post-selected bipartite systems with $N$ photons at local system $A$ and $M-N$ photons at local system $B$.

A more specific analysis is necessary to determine the quantum state fidelity at which the tighter bound of Eq.(\eqref{eq::ent_criterion_optimized}) will be exceeded. Here, we have to take into account the pattern class distribution and the associated values of $D_{\boldvec{p}}$.
As shown above, the ideal state has $D_{\boldvec{p}}=1/3$ and hence the quantity $F_{\boldvec{n}}-D_{\boldvec{p}}+F_{K}$ has the value $5/3$ for the ideal component $\ket{\phi_{2_{A},2_{B}}}$, which contributes a fraction of $F_{\text{state}}$ to the statistics.
Here, we also need to consider the possible contributions to $D_{\boldvec{p}}$ by the error statistics.
However, the error statistics may contribute a different value of $D_{\boldvec{p}}$. The most conservative assumption that the errors originates from the smallest complementary pattern class $(\mathcal{E}_{1010},\mathcal{E}_{0101})$ would assign the maximal value $D_{\boldvec{p}}=1/2$ to all of the errors,
and the quantity $F_{\boldvec{n}}-D_{\boldvec{p}}+F_{K}$ has the lower bound $-1/2$, which contributes a fraction of $1-F_{\text{state}}$ to the statistics.
This resulting in a bound of
\begin{equation}
\min(F_{\boldvec{n}} - D_{\boldvec{p}} + F_K) = \frac{13}{6} F_{\text{state}} - \frac{1}{2}.
\end{equation}
From this bound, we find that a quantum state fidelity of $F_{\text{state}}>9/13$ is sufficient for an experimental verification of entanglement using the criterion of Eq. \eqref{eq::corr_fidelity_upper_bound}. In terms of the numerical value, this is a quantum state fidelity of about $69.23$ percent.

Finally, it may also be worth noting that the expectation value of the witness operator $\hat{W}$ for a completely random state $\hat{\rho}=\hat{1}/100$ can be obtained from Fig. \ref{fig::opt_sep_2A2B_corr_fidelity} by counting the number of outcomes in the red dashed squares. The results are $0.06$ for the fidelity $F_{\boldvec{n}}$, $0.26$ for the fidelity $F_K$, and $0.06$ for the pattern class defect $D_{\boldvec{p}}$. Therefore, the expectation value of $\hat{W}$ for a completely random state is $-0.74$. For the ideal state, the expectation value of the witness is $2/3 \approx 0.67$. For a statistical mixture of the random state and the ideal state given by
\begin{equation}
\hat{\rho} = p \projector{\phi_{2_{A},2_{B}}} + (1-p) \frac{1}{100} \id,
\end{equation}
the expectation value of the entanglement witness $\hat{W}$ is therefore given by $1.41 p - 0.74$, which exceeds zero for $p > 0.525$ . We can therefore conclude that the experimental entanglement criterion given by Eq. (\eqref{eq::ent_criterion_optimized}) and the corresponding entanglement witness $\hat{W}$ can detect entanglement even when about 45 percent of the state are completely randomized.

\section{Conclusion}
\label{sec::conclusions}

In conclusion, we have presented a practical method of evaluating the entanglement between two local multi-mode systems that can be generated and scaled up easily if sufficiently reliable single photon sources and linear optics circuits are available.
As we have shown in section \ref{sec::ent_generation}, the coherent distribution of $M$ photons in $M$ modes to two different locations by beam-splitting naturally results in large amounts of entanglement between the two $M$-mode systems.
However, it will be difficult to avoid experimental imperfections as the number of photons and modes increases.
It is therefore essential to identify experimental criteria that can verify the entanglement through measurements of the photon statistics that characterize the state.
This can be done by measuring two sets of non-commuting photon number distributions that are related to each other by a linear optics transformation.
Here we argue that the most suitable mode transformation for this purpose is the discrete Fourier transformation (DFT).
The highly symmetric properties of the DFT allows us to formulate the \emph{mode shift rule} of DFTs, which describes a well-defined relation between quantum coherences between different photon number distributions in the input of the DFT and the experimentally observable output photon number distributions of the DFT.
We can then organize the correlated measurement results obtained with and without local DFTs in such a way that we can identify correlations exceeding the limits that apply to all separable states.
In section \ref{sec::ent_crit} and \ref{sec::tighter_bound}, we have presented specific entanglement witnesses that can be used to verify the entanglement generated by beam-splitting multiple single photon inputs.

The present work provides the necessary tools for the characterization and control of entanglement between multi-mode systems with high photon numbers.
It thus opens up new possibilities for the generation and application of quantum states in large scale systems. Since much effort is currently made to increase the size of optical quantum circuits, both in terms of the number of modes and in the number of photons, the present work may serve as a guide to the new territory pioneered by these technological advances. The concepts developed above should provide helpful examples of how to address the increasing complexity of large scale quantum systems. In particular, the DFT and its mode shift rule appear to be an important and highly versatile element in the quantum optical toolbox for the development of future quantum technologies. We therefore hope that the present work provides a useful starting point for a more detailed exploration of the wide variety of new experimental possibilities that are beginning to emerge as a result of ongoing research in the field.

\acknowledgments
This work was supported by JSPS KAKENHI Grant Number 26220712.

\bigskip
%
\bibliographystyle{iopart-num}
\bibliography{multiph_ent_from_single_ph}

\end{document}